\documentstyle[aps,prb,preprint]{revtex}
\begin{document}
\title{
Two--magnon scattering and the spin--phonon interaction beyond the
adiabatic approximation}
\bigskip
\author{Matthew J. Reilly and A. G. Rojo}
\address{Physics Department, The University of
Michigan,
Ann Arbor, MI 48109-1120
}

\maketitle

\begin{abstract}

We consider a model of Raman scattering for a two--dimensional $S=1/2$
Heisenberg Anti-Ferromagnet
which  includes a {\it dynamical} spin--phonon
interaction. We observe a
broadening of the
line shape due to increased coupling with excited high--energy spin states.
Our
results
are close to a model of random, static exchange interactions,
first introduced in this context
by Haas {\it et al.} [J. Appl. Phys. {\bf 75}, 6340, (1994)],
 which, when
extended to large
numbers of spins,
explains experiments in the parent insulating compounds of
high-$T_c$ superconductors.
\end{abstract}

\section{Introduction}

Since the discovery
of high--$T_c$ superconductors\cite{bendorz}, the $S=1/2$ quantum
Heisenberg model  has received considerable attention.
This is largely due to well-accepted experimental evidence
 that
 suggests that these compounds can be described\cite{manousakis} by a
 two--dimensional (2D) doped Heisenberg
 model
 for spin $S=1/2$.

 One of the experimental techniques used to study the excitations of
 these systems is Raman scattering.
 This technique was intensely used in the past for different
 antiferromagnets of spin $S\ge 1$.
 For such systems, the main features of the
line shape were explained by
Parkinson\cite{parkinson}, who used a  spin--wave theory to
account for the photon--two-magnon process involved in the
Raman scattering.  Very good agreement\cite{fleury} was found for
K$_2$NiF$_4$, which
is well described by a spin $S=1$ Heisenberg antiferromagnet.

The results of Raman experiments in the
parent insulating compounds of high--$T_c$ superconductors,
La$_2$CuO$_{4}$ and YBa$_2$Cu$_3$O$_{6.2}$, show some qualitative
differences with the line shape of previously studied
antiferromagnets\cite{sugai,sulewski,lyons}.
As in the case of K$_2$NiF$_4$, the line shape is centered at
an energy corresponding to a spin--pair excitation. However, in contrast
with the case of spin $S=1$, the linewidth is very broad, and the spectrum
has
a very long tail that extends beyond  the energy of four magnon excitations.
Moreover, while the dominant contribution to scattering is in the so--called
$B_{1g}$ channel, there is also a significant contribution in
the nominally forbidden $A_{1g}$ configuration\cite{lyons}.

Since for lower spins the quantum fluctuations are larger,
some theoretical analysis beyond the mean--field spin--wave
approximation has been attempted
to explain the broad line shape.
Numerical diagonalizations in a $4 \times 4$ lattice shows very
little structure for the $B_{1g}$ channel: essentially three peaks;
a dominant two--magnon peak and two peaks identifiable as four--magnon
excitations\cite{GBC}. This calculation gives vanishing line shape for the
$A_{1g}$ channel.
Although  the structure of the line shape is clearly different
 from the one observed in the experiment, the first
 three moments of both lines are in good agreement. The first three moments
 of the distribution have been calculated in good agreement to experiment by
 Singh {\it et al}.
 using cumulant expansions on a Heisenberg model with diagonal
 next--nearest--neighbor couplings\cite{singh}.
 Canali and Girvin\cite{CG} used the Dyson--Maleev transformation
 taking into account processes of up to four magnons, and presented convincing
 evidence that the line shape cannot be explained by
 quantum fluctuations alone.
 Raman scattering Hamiltonians with a four--site
 exchange\cite{honda} term have been proposed, these increase
  the relative weight of the four--magnon
 scattering, but does not explain the  broadening of the peas.
 This is also
 consistent with
 the work of Sugai\cite{sugai1}, and Roger and Delrieu\cite{roger}.
 Also, recent studies of spin--pair excitations in a spin $S=1$ system,
 NiPS$_3$, show a similar linewidth as those observed
 in the cuprates\cite{merlin}.

 From the above considerations, one concludes that it is necessary
 to invoke an additional process. It was emphasized by
 Nori {\it et al}.\cite{nori}
 that the  spin--phonon interaction
 can be responsible for the broad linewidth, and for the finite cross--section
 in the otherwise absent $A_{1g}$ channel.
They supported their arguments by computing the Raman cross section
 in finite lattices, using a Heisenberg model with random {\it static}
 exchange integrals, and averaging over configurations with equal
 weight.
 This calculation is in the spirit of the adiabatic or Born--Oppenheimer
 approximation\cite{ziman}, that neglects the
 fluctuations of the phonon field.

Adding phonons to the 2D $S=1/2$ Heisenberg Hamiltonian (HAF)
 provides a mechanism\cite{nori} for the otherwise
forbidden $A_{1g}$
scattering, as well as allowing a coupling between the ground and excited
spin states.
In a typical HAF, the
phonon frequency\cite{nori}
is about a third of
the interaction constant $J$. The first excited state lies
 at $3J$,  so there is an order of magnitude energy
difference  between spin and phonon excitations.
It was argued\cite{nori} that the separation of energy scales justifies the
adiabatic phonon approximation.

 In the present work we consider the Raman scattering for  a Heisenberg
 model with spin--phonon interaction, including the effects of the phonon
 dynamics. We solve for the exact ground state of a system in which
 the vibrational degrees of freedom are allowed a finite number
 of excitations, an approximation valid for small but finite
 phonon frequencies.
To the best of our knowledge, this is the first calculation
of exact diagonalizations for a dynamic spin--phonon system.
Some work including phonon dynamics beyond the adiabatic
approximation exists in
the context of one--dimensional  Peierls systems. Fradkin and
Hirsh\cite{fradkin} studied the electron--phonon interaction
 using quantum Monte Carlo
simulations, and Proetto and Falicov\cite{proetto}
solved the case of two sites and one phonon.

This paper is organized
as follows. In Sec. II we present our Hamiltonian and scattering operator
 and discuss the rationale for our truncated--phonon model.
  In Sec. III we present computational results and provide theoretical support
for these results.
We also present an
alternative approach to including phonons and compare it to our model.
Sec. IV is devoted to conclusions.

\section{Model and Procedure}
We  study a Heisenberg model with spin--phonon interaction described by
the following Hamiltonian:
\begin{equation}
H = \sum_{<ij>} \left\{ \left[J - \alpha (a^{\dagger}_{ij}+a_{ij})\right]
\vec S_{i} \cdot \vec S_{j}
+ \omega_0 a^{\dagger}_{ij}a_{ij} \right \}
\label{H} \,
\end{equation}
where $\vec S_{i}$ are spin $1/2$ operators, $a^{\dagger}_{ij}
(a_{ij})$
is a creation (annihilation)
operator for an Einstein phonon of frequency $\omega_0$, and the simbol
$``ij"$
refers to a link of a square lattice.
In $J_{ij} = J - \delta J_{ij}$ we include the coupling
 between the phonons and the spin degrees of freedom
 through a ``displacement" operator
 $\hat x_{ij} = \sqrt{\frac{\hbar}{2m\omega_0}}
(a^{\dagger}_{ij}+a_{ij})$.
 Our
parameter $\alpha $ in
 Eq.(\ref{H}) is proportional to the  spin--phonon  coupling constant
$\lambda=\alpha \sqrt{{2m\omega_0}/{\hbar}}$ relating the change in
the exchange integral $\delta J_{ij}$
 with the displacement: $\delta J_{ij} =\lambda \hat x_{ij} = \alpha
 ( a_{ij}^{\dagger} + a_{ij})$.

Due to computational  limitations, we need to restrict ourselves to a small
number ($6$ and $8$)
of spins.  To have these as two dimensional as possible, we
placed them in
non-periodic ladder type structures as shown in  Fig. \ref{Figure1}.

We use Einstein phonons to simplify the model, and we consider the
highest--energy
 phonons as they are the closest in energy to the magnon excitations
and will thus have the greatest coupling to the magnons.
The occupation number of the phonon degrees of freedom at each
link is in principle unrestricted; the number of phonons ranges
from zero to infinity. This makes the problem intractable from the point
of view of exact diagonalizations,
since the resulting Hilbert space is infinite.
In order to overcome this limitation, we restrict the possible
number of phonons at each link
 by imposing the
 condition $(a^{\dagger})^{n} = 0$. Computational resources limit us to
using  $n = 2,3$.
This approximation maps the phonon degree of freedom into a two-- or
three--state system
at each link.
The truncation of the phononic occupation states implies that the
 variations of $J_{ij}$ due to quantum fluctuations
are bounded. For example, for $n=2$,
the maximum displacements for a given link is given by the ``coherent"
states
$|\psi_{ij}^{\pm}\rangle=2^{-1/2}(1\pm a^{\dagger}_{ij})|0\rangle$,
and $\delta J_{ij}= \alpha \langle \psi_{ij}^{\pm}|
(a^{\dagger}_{ij}+a_{ij})|\psi_{ij}^{\pm}\rangle=\pm\alpha$.
Quantum fluctuations themselves are certainly limited by
our
truncation, but since they are in general small for small $\omega_0$,
 our approximation will account for the relevant dynamics in the
regime $\delta J_{ij} /J < 1$ and $\omega_0 /J <1$.

We are  interested in the  Raman scattering intensity $
I_R(\omega)$, given by:
\begin{equation}
I _R(\omega)=\sum_{\nu}\: |\langle \nu|\hat R|0\rangle|^{2}
\:\delta(\omega-
E_\nu+E_{o})
\label{sigmar}  \, ,
\end{equation}
where $|\nu\rangle$ are the eigenstates  and $E_\nu $ the eigenvalues
of $H$.
We
compute $ I _R(\omega)$
 by using the partial
 fraction expansion method of Gagliano and Balseiro\cite{GB}.
The relevant operator $\hat R$ in Eq.(\ref{sigmar})
depends on the different configurations of the Raman
experiment. In general\cite{parkinson}:
\begin {equation}
\hat R = \sum_{<ij>} (\vec E_{\rm inc}\cdot\hat x_{ij})
(\vec E_{\rm scat}\cdot\hat x_{ij})
\vec S_{i}\cdot \vec S_{j}
\label{A} \, ,
\end{equation}
with $\vec E_{\rm inc}$ and $\vec E_{\rm scat}$ refer to the polarization
of
the incoming and scattered photon  respectively, and $\hat x_{ij}$
are unit vectors in the near--neighbor directions of the lattice.
 For square lattices, two common
scattering  configurations
are the so--called $B_{1g}$  and $A_{1g}$ symmetries.
We let  the lattice axis lie along the $x$
and $y$ directions, and we define  $x'$ and $y' $ the directions along
the lattice diagonals ($\hat{x'} = \frac{\hat{y}+\hat{x}}{\
\sqrt{2}}$
and $\hat{y'} = \frac{\hat{y}-\hat{x}}{\sqrt{2}}$),
then $B_{1g}$ corresponds to
 $\vec E_{\rm scat} \parallel \hat{x'}$ and $\vec  E_{\rm inc} \parallel
\hat{y'}$,
 whereas $A_{1g}$ corresponds to both the incident  and scattered  photon
polarized
 in the same direction: $\vec E_{inc}\parallel \vec E_{\rm scat} \parallel
\hat{x'}$.

For the case of the ``pure" Heisenberg Hamiltonian
(no phonons, or
$\alpha =\omega_0=0$),
the
 absorption
corresponding to the $A_{1g}$ symmetry  is zero at any non-zero frequency,
since the $A_{1g}$ operator is  proportional to  the Hamiltonian.
We show in the next section how the inclusion of the phonon dynamics
gives rise to a finite absorption in this channel, and
argue that a spin--phonon interaction accounts for the
main features observed in the Raman experiments in the undoped
copper oxides.
Also, we compare the results obtained in the dynamical model
with that of the Heisenberg model with random (static)
exchange integrals, which   was studied in this context by Nori
{\it et al}.\cite{nori}

\section {results and discussion}

In this section we present
numerical  results for systems of $6$ and $8$ spins\cite{hilb}
corresponding to the geometry shown in Figure \ref{Figure1}.
We first discuss the $A_{1g}$ symmetry. In this case, and in
the absence of lattice distortions, the Raman operator
commutes with the Heisenberg Hamiltonian.
No line shape is observed in this case:
$I_R(\omega) \sim \delta (\omega)$.
It was first pointed out by Nori {\it et al}.\cite{nori}
that the presence of {\it static}
disorder in the exchange integrals $J_{ij}$ changes the value of the
commutator, and
can give rise to a finite $A_{1g}$ signal quite similar to experiments.
It was  argued by Nori {\it et al}.\cite{nori} that the spin--phonon coupling
produces disorder in the values of $J_{ij}$,
 in the limit where the vibrational
motions are much slower than the spin degrees of
freedom. Here, we are interested  in how this limit is achieved
in a system that---up to boundary
effects---is translationally invariant, and includes the
{\it dynamics} of the spin--phonon coupling.
Consider our Hamiltonian Eq.(\ref{H}) and the operator $\hat R$ for
the $A_{1g}$ symmetry.
For $\omega_0 \neq 0$, $[H,\hat R] \neq 0 $ and we expect a finite
line shape. In Fig. \ref{Figure2} we show the  $I(\omega)$  obtained
for  $\alpha=0.1J$ and  $\omega_0=0.05J$.
Some of the qualitative features of the experimental line shape  are already
present in these finite--size systems: there is a broad
line shape of width $\sim 2J$ with a maximum at the two--magnon
energy. Note that for our ladder geometries,
since most sites have coordination number $3$, the two--magnon
energy in the Ising limit is located at $2J$,  instead
of the corresponding $3J$ of the square lattice.
Due to the finite size of the lattice, the line shape consists
of a series of peaks that are centered at the unperturbed
energies, which are indicated by arrows in  Fig. \ref{Figure2}. These
unperturbed energies
correspond to the manifold of two--and multi--magnon states.
In turn, these ``internal" peaks have a finite width that
is due to the coupling with the phonons.
The energy scale dominating this width is $\alpha$.

In order to test the validity of the Hilbert space truncation,
we have increased the number of allowed excitations per phonon by one, with
no qualitative changes in the line shape (see Fig. \ref{Figure3})

It is interesting to consider the following  two
limiting cases: $(a)$  vanishing spin-phonon coupling $\alpha \rightarrow 0$
 for  finite phonon frequency $\omega_0$,
and $(b)$ vanishingi
phonon frequency $\omega_0 \rightarrow 0$ for finite $\alpha$.
In both limits the  $A_{1g}$ line shape vanishes.
In  $(a)$, the spins are uncoupled from the phonons,
and the line shape vanishes because in this
limit $[H,\hat R]=0$.
 Figure \ref{Figure4} illustrates  how the line shape
vanishes and shifts to lower energies as $\alpha \rightarrow 0$.
In limit $(b)$, the phonons are ``static", but remain coupled
to the spins, because $\alpha \neq 0$. Since the phonons  have
no dynamics, the eigenstates can be written  in the  form
\begin{equation}
|\psi _\nu (\omega _0=0)\rangle =
|\psi _\nu \{n\}\rangle \otimes |\{n\}\rangle \, ,
\end{equation}
where $|\{n\}\rangle$ is a static configuration of displacements,
in such a way that, for each link $ij$, one has
$(a^{\dagger}_{ij}+a_{ij}) |\{n\}\rangle=
\eta_{ij}({\{n\}}) |\{n\}\rangle$, with $\eta_{ij}(\{n\})$ a $c$--number.
The state $|\psi _\nu \{n\}\rangle$ involves spin degrees of freedom
only, and is an eigenstate of the following Hamiltonian
\begin{equation}
H\{n\}  =  \sum_{<ij>}
\left[J -\alpha \eta_{ij}\left(\left\{n\right\}\right)\right]
 \vec S_{i} \cdot \vec S_{j}
   \, .
  \end{equation}

  For $\omega_0=0$, the
  problem is that of an {\it annealed} configuration of displacements:
  for each configuration $\{n\}$, the couplings $J_{ij}$ are different,
  and one has to solve for the spin dynamics in the presence of
  this distribution of couplings. The complete spectrum is obtained
  by solving $N_p$
  decoupled eigenvalue problems, with  $N_p$ being
  the phononic Hilbert space dimension. For our case of one
  excitation per link, $N_p = 2^{N_{\rm links}}$, with
  ${N_{\rm links}}$ being the number of links of the lattice.
What distinguishes this from a quenched disorder is the fact that
  the ground state of this problem corresponds to the lowest energy
  state of the spin system in an {\it ordered} background of
  couplings.  Since this is also an eigenstate of $\hat R$ for
  the $A_{1g}$ symmetry, the line shape will be zero. In Fig. \ref{Figure5}
  we show  how the line shape vanishes as $\omega_0$ decreases to
  zero.
Note that from this analysis one concludes that,
even in the case of  $\omega_0=0$, the line shape
is non--zero at finite temperature. This is due to the presence of
thermally excited states, which are eigenstates of a Hamiltonian
that does not commute with $\hat R$.
The right column
of Figure \ref{Figure5} shows the rescaled results, which illustrate
 the line shape approaching a limiting function of $\omega$ whose
overall amplitude
vanishes as $\omega_0 \rightarrow 0$. The main features of this
limiting function
should be accounted for by a model of static disordered couplings.
In previous work in the context of one--dimensional Peierls
systems\cite{wilkins_igor},  it is argued that
in the limit of small phonon frequencies,
the quantum lattice fluctuations can be  modeled by a static,
random Gaussian potential with zero mean.
A model in which the exchange integrals are taken from a random configuration
of couplings was also studied by Nori {\it et al}\cite{nori}.
We can test this hypothesis in
our dynamical
model.  Results of the comparison between the dynamical and a static
disordered system are shown in Fig. \ref{Figure6}.
Note that
both the amplitude
and the position of the peaks are in very good mutual agreement.
The  scattering peaks from  the two models have their energies scaled with
 respect
 to each other.  After they are rescaled, the
 location and magnitude of
 the peaks are very close for both models (see Fig. \ref{Figure6}).
 The basis for this rescaling of the horizontal axis is the following.
   For our
    dynamic phonon model, the
    minimum
    energy state corresponds to the lattice ``maximally compressed":
    $\eta_{ij}=-\langle x \rangle_0$.
     The tendency of the system to compress has the
     effect of
     renormalizing the
     interaction coefficient $J$, since the displacement $x_{
     ij}$
     will tend to
     be a constant, and therefore $J' = J + \alpha \langle x \rangle_0$.  The
      average displacement will be
     greater
     as the possible number of excitations increases;   this
     explains a shift in
     energy between $1$ and $2$ excitations as well.

In the model with
static disorder, the
phonon energy goes to zero, but  the fluctuations are still present.
In the treatment of Nori {\it et al}.\cite{nori}
the argument
used  is that the phonon energy is small compared to that of the spin
excitations,
 and thus can be neglected.
  However,
  we have shown that for $\omega _0 $ strictly zero the line shape vanishes.
   At zero temperature,
   the disordered model should be compared with the dynamical model
    at infinitesimal $\omega_0$. We prove this statement
    by
    using perturbation theory in $\omega_0$. Assume an ordering of the states
     for $\alpha =\omega_0 =0$, and let us label those states by the index
      $\nu$. If $\alpha \ll J$ and $\omega_0 =0$,
    each state $|\psi_\nu \rangle $ splits in  a manifold
    $|\psi_\nu\{n\} \rangle $ of almost degenerate states of energies
     $E_\nu\{n\}$, that correspond to the ``disordered" configurations.
      If we now turn on $\omega_0$, different states are going to mix,
      in  such a way that the ground state can be written as
      \begin{equation} |\psi_0\rangle \simeq |\psi_0\{n\}_0 \rangle +
 \omega_0\sum_{\{n\}\neq\{n\}_0} c(\{n\})|\psi_0\{n\}\rangle
  \otimes |\{n\}\rangle \, ,
 \end{equation}
where $\{n\}_0$ is the ordered configuration of couplings.
A similar expansion can be used for the excited states.
If we use the fact that $A_{1g}$ $\hat R$ acts only on the spin degrees of
freedom, and that different states $ |\{n\}\rangle$ are orthogonal,
to lowest order in $\omega_0$, $I(\omega)$ will be given by
\begin{equation}
I(\omega)\simeq \omega_0^2\sum _{\{n\}} |c(\{n\})|^2
\sum_\nu |\langle \psi_\nu\{n\}|\hat R |\psi_0\{n\}\rangle|^2
\delta(\omega -E_\nu\{n\}+E_0\{n\}) \, .
\end{equation}

We have found
computationally a falloff of $I(\omega)$ proportional to $\omega_0^2$.

Now, if we keep the assumption of $\alpha \ll J$, the states corresponding
to each manifold are almost degenerate and we can approximate
$c(\{n\}) \sim (M-1)^{-1/2}\sim {M}^{-1/2}$,
with $M$ the
number of configurations  $\{n\}$. Therefore, in this limit $I(\omega)$
is given by an average over configurations with {\it equal} weight.

We now
present results
for the $B_{1g}$ configuration. In Fig. \ref{Figure7} we compare
the
$B_{1g}$ scattering from eight $S=1/2$ Heisenberg spins
for three different cases: bare HAF, with phonon interactions,
and with static disorder.

We have argued that peaks at higher  excitations will eventually merge
into one broad peak
as the lattice
size increases.  To show that this is  the case, we have calculated
the $A_{1g}$ and
$B_{1g}$ peaks
for a $3\times 4$ non--periodic lattice with static disorder.  The
results are shown in Fig. \ref{Figure8}.
These are in good qualitative agreement
with experiments,
except that the experimental scattering is centered at $\omega \sim 3J$
instead of $2J$ in our case.
  This is due to finite size effects and to the lattice being non--periodic.

The  model with static
disorder, when extended to larger numbers of spins than can be obtained
with the
dynamic phonon model,  begins to duplicate experiments.  The agreement
between the static and
dynamic phonon
models  suggests that the dynamic phonon model will also agree with
experiments.

\section{Conclusions}
In this paper we have presented a model of phonon-magnon interaction with
a truncated phonon space.
We discussed how this model can explain the otherwise
forbidden $A_{1g}$
scattering and showed that it does in fact give rise to $A_{1g}$
scattering. We discussed how coupling to spin excitations leads
to a broadening
of the line shape. We have proven that phonons can be modeled by
static disorder and compared
the results obtained by using quenched phonons to our dynamical model.
 We showed that the amplitude of the $A_{1g}$ scattering should fall off
as $\omega_0^2$.
We showed that both the $A_{1g}$ and $B_{1g}$ scattering is broadened.
 In conclusion,
the broadening is due
to coupling to spin excitation states through the phonon interaction.

\section{acknowledgements}

We would like to
thank Roberto Merlin and Jim Allen for helpful discussions, and
Franco Nori for helpful discussions and a careful reading of this manuscript.

\begin{figure}
\caption
{ Ladder geometries used in our numerical computations.
Each dot
represents a spin $1/2$.  Each solid line connects nearest-neighboring
spins, and
represents a Heisenberg Anti-Ferromagnetic coupling modulated by an
Einstein phonon. There is one phonon per coupling.
\label{Figure1}
 }
\end{figure}

\begin{figure}
  \caption{
  Raman cross section $I(\omega)$
  versus $\omega /J$ for $A_{1g}$ scattering. The
  parameters of the dynamical Hamiltonian are $\alpha  = 0.1J$,
  $\omega_0 = 0.05J$, and the number of phonon excitations per link, $n$,
   is $2$. The vertical scale is arbitrary.
  The arrows  are located at the eigenenergies of the model without phonons
  corresponding to eigenstates of total spin $S=0$.
  \label{Figure2}
  }
   \end{figure}

\begin{figure}
\caption{
Comparison between one--and two--phonon (lighter curve) excitations
per link,
for the Raman line shape $I(\omega)$ in the $A_{1g}$ symmetry.
The parameters are
$\alpha=0.1J$, and $\omega_0 = 0.05J$, for a ladder of $6$ spins.
\label{Figure3}
 }
 \end{figure}

\begin{figure}
\caption{
Raman line shape for the $A_{1g}$ symmetry.
Here we consider $6$  spins (including $1$
excitation per link), and $\omega_0 = 0.05J$. The strength of the
spin-phonon coupling is varied: (a) $\alpha = 0.2J$, (b) $\alpha = 0.1J$
(c) $\alpha = 0.05J$, (d) $\alpha = 0.025J$.
The vertical scale is arbitrary but is the same for the different
values of $\alpha$.
 }
\label{Figure4}
\end{figure}

\begin{figure}
 \caption{
 Raman spectrum versus $\omega/J$ for the $A_{1g}$  symmetry.
 These results correspond to $6$  spins coupled to one phonon
 excitation  and $\alpha = 0.1J$. (a)-(c) are all displayed on the same scale,
 (d)-(f) have been rescaled.
 $\omega_0 = 0.3$ for (a) and (d), $\omega_0 = 0.1$ for (b) and (e), and
 $\omega_0 = 0.025$ for (c) and (f).
 \label{Figure5}
  }
\end{figure}

\begin{figure}
\caption{
Raman spectra for the $A_{1g}$ symmetry for $8$ spins
coupled to one phonon excitation
and also scattering from
an average
over configurations of disorder. (a) $\alpha = 0.1J$, $\omega_0 = 0.05J$,
and disorder
chosen from a Gaussian with variance $\sigma= 0.2J$. (b) $\alpha = 0.3J$,
$\omega_0 = 0.3J$,
and disorder chosen from a Gaussian with variance $\sigma= 0.5J$.
 \label{Figure6}
 }
 \end{figure}

\begin{figure}
\caption
{
Raman scattering for the $B_{1g}$ symmetry using
(a) the bare Heisenberg Hamiltonian ($\alpha= 0$),
(b) with dynamical phonons
$\alpha =0.1J$,
$\omega_0 = 0.05J$, (c) with dynamical phonons $\alpha = 0.3J$,
$\omega_0 = 0.3J$. The  two  bottom curves correspond to static
Gaussian disorder of variance $\sigma= 0.2J$ (d), and $\sigma= 0.5J$ (e).
\label{Figure7}
 }
\end{figure}

\begin{figure}
  \caption{
Raman scattering
from a $3\times 4$ non-periodic lattice in the $A_{1g}$ and $B_{1g}$
geometries.
The spin interactions
$J_{ij}$ have static Gaussian disorder of variance $\sigma = 0.5$
    \label{Figure8}
  }
   \end{figure}


\begin{references}
\bibitem{hilb} The Hilbert space dimension is $ \left(
\begin{array}{c}
{\textstyle N}\\[-10pt]{{N}/{2}}\end{array}\right
) n_{e}^{n_{b}}$, where $N$ is
the number of
spins, $n_{e}$ is the number of allowed phonon states, and $n_{b}$
is the number of
nearest-neighbor pairs.  For $6$ spins and $2$ ($3$) phonon states
in a ladder configuration,
this gives a
Hilbert dimension of $2560$ ($43740$).  For $8$ spins, $2$ ($3$) excitations,
the dimension becomes $71680$ ($4133430$).
\end{references}
\end{document}